\documentclass[aps,prl,twocolumn,superscriptaddress]{revtex4-1}
\usepackage{graphicx}
\usepackage{epsfig}
\usepackage{longtable}
\usepackage{amsmath}
\usepackage{amssymb}
\usepackage{color}

\begin{document}

\title{Functional Mn--Mg$_k$ cation complexes in GaN featured by Raman spectroscopy}

\author{T.~Devillers}
\email{thibaut.devillers@jku.at}
\affiliation{Institut f\"ur Halbleiter-und-Festk\"orperphysik, Johannes Kepler University, Altenbergerstr. 69, A-4040 Linz, Austria}

\author{D.~M.~G.~Leite}
\affiliation{Instituto de F\'isica e Qu\'imica, Universidade Federal de Itajub\'a, 37500-903, Itajub\'a--MG, Brazil}
\affiliation{S\~ao Paulo State University, · Department of Physics,  Bauru--SP, Brazil}

\author{J.~H.~Dias~da~Silva}
\affiliation{S\~ao Paulo State University, · Department of Physics,  Bauru--SP, Brazil}

\author{A.~Bonanni}
\email{alberta.bonanni@jku.at}
\affiliation{Institut f\"ur Halbleiter-und-Festk\"orperphysik, Johannes Kepler University, Altenbergerstr. 69, A-4040 Linz, Austria}

\date{\today}

\begin{abstract}

The evolution of the optical branch in the Raman spectra of (Ga,Mn)N:Mg epitaxial layers as a function of the Mn and Mg concentrations, reveals the interplay between the two dopants. We demonstrate that the various Mn-Mg-induced vibrational modes can be understood in the picture of functional Mn--Mg$_k$ complexes formed when substitutional Mn cations are bound to $k$ substitutional Mg through nitrogen atoms, the number of ligands $k$ being driven by the ratio between the Mg and the Mn concentrations.

\end{abstract}

\maketitle

Nitride semiconductors have increasingly attracted attention over the last decades, particularly due to their remarkably broad spectrum of technologically relevant applications\,\cite{MorkocBook2008}. While nitride-based compounds undoubtedly prevail as building-bocks for visible and ultraviolet solid-state optoelectronic and high power electronic devices, the fields of spintronics\,\cite{DietlNatMat2010}, thermoelectricity\,\cite{LuSST2013}, and terahertz electronics~\cite{NepalAPEX2013} are among the ones where nitrides are currently giving rise to a growing interest for prospective functionalities. The realization of a new generation of nitride-based devices requires a profound understanding and control of the doping mechanisms in these materials. In this context, it is becoming increasingly clear that the traditional picture of isolated impurities is not sufficient to properly illustrate the properties of these materials systems, particularly when two or more dopant species are involved and/or in regime of high concentration. We have recently demonstrated how the formation of Mn--Mg$_k$ complexes in (Ga,Mn)N:Mg substantially affects the behaviour of each dopant, resulting in remarkable and unforeseen material properties \cite{Devillers2012}. The interplay between the dopants, but also with the host lattice must be understood in order to ensure a controlled and effective functionality of these nanoobjects in innovative devices. 
Raman spectroscopy is particularly suitable to investigate impurities and impurity complexes in semiconductors \cite{WagnerASS1991} due to its sensitivity to variations in the structural and electronic properties of the material under study. In the case of doping of nitrides with Mg, this technique was decisive to identify the Mg--H complexes\,\cite{GotzAPL1996b,CuscoJAP2012} responsible for the low efficiency of the Mg activation, and for the consequent poor $p$-type conductivity especially in GaN:Mg \cite{Brandt1994,Harima1999,Kaschner1999}.
While Mg is the most relevant $p$-type dopant in GaN, and is employed in the majority of optoelectronic devices based on GaN, Mn gained interest lately for its potential in spintronics, and particularly in the perspective of turning $p$-type Mn-doped GaN ($i.e.$ (Ga,Mn)N:Mg) into a ferromagnetic dilute magnetic semiconductor with a high Curie transition temperature \cite{Dietl2001}. Moreover, the recent demonstration of unexpected (infrared) optical and magnetic responses in (Ga,Mn)N:Mg \cite{Devillers2012}, has opened  appealing fundamental and technological perspectives for this system.  

In this letter, we  apply Raman spectroscopy at room temperature in order to investigate the local environment of Mn and Mg in (Ga,Mn)N:Mg.
The samples consist of a 1\,$\mu$m thick GaN buffer layer grown on $c$-sapphire (0001), on which a 600~nm layer of GaN doped with Mn and/or Mg is deposited. The samples are fabricated by metalorganic vapor phase epitaxy (MOVPE) according to a procedure detailed elsewhere\,\cite{BonanniPRB2011,Devillers2012}. Prior to Raman spectroscopy, the structural, magnetic and chemical properties of the samples are extensively investigated~\cite{Devillers2012} and in particular the crystalline structure is analysed by synchrotron high-resolution x-ray diffraction and high-resolution transmission electron microscopy, showing that all samples are single-crystalline and do not include secondary phases. Furthermore, synchrotron x-ray absorption fine structure (EXAFS) studies demonstrate that at least 95\% of the Mn ions substitutes Ga cations\,\cite{Stefanowicz2010}, and we have reported that in the case of Mn-Mg codoping, both Mn and Mg occupy Ga substitutional sites, and Mg tends to stay close to Mn, separated by one N atom, to form Mn--Mg$_k$ cation complexes, where $k$ is the number of Mg ions in the first cation shell of a single Mn~\cite{Devillers2012}. The Mn and Mg concentrations $x_\mathrm{Mn}$ and $x_\mathrm{Mg}$ are extracted from secondary ion mass spectroscopy (SIMS) and lie between 0 and 1\%. The Raman spectra are acquired with a Jobin-Yvon LabRAM HTS confocal micro-Raman setup equipped with 532~nm, 633~nm and 785~nm lasers with a spectral resolution of 1~cm$^{-1}$. All spectra presented here are acquired using the 532~nm beam, in $Z(X-)\overline{Z}$ configuration, $i.e.$ at normal incidence and emergence, with a linearly polarized  incident beam, and  unpolarized scatterd beam. After the subtraction of a linear background, all the spectra are renormalized with respect to the intensity of the E$_2^H$ peak. Particular attention is given to the optical branch of GaN, especially to the energy range between 500~cm$^{-1}$ and 800~cm$^{-1}$.

\begin{figure}[ht]
	\begin{center}
	\includegraphics[width=0.95\linewidth]{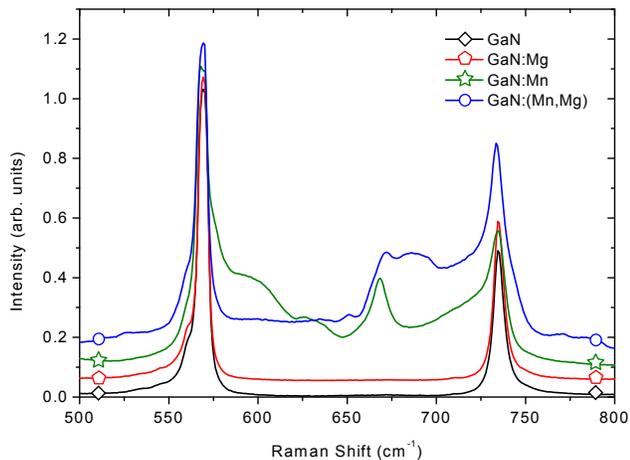}
	\caption{Raman spectra of: nominally undoped GaN, GaN:Mg, (Ga,Mn)N, and (Ga,Mn)N:Mg. To clarity, an $y$-offset of 0.05 has been added between the curves.}
	\label{fig:fig1}
	\end{center}
\end{figure}

In Fig.\,\ref{fig:fig1}, the Raman spectra for four representative samples are reported, namely GaN, GaN:Mn ($x_\mathrm{Mn}$=0.75\%), GaN:Mg ($x_\mathrm{Mg}$=0.25\%), and GaN:(Mn,Mg) ($x_\mathrm{Mn}$=0.25\% , $x_\mathrm{Mg}$=0.45\%).

In the present geometry, the optical branch of the GaN spectra is dominated by two modes: the transverse E$_2^H$ mode at 568~cm$^{-1}$ and the longitudinal A$_1$(LO) at 733~cm$^{-1}$. The A$_1$(TO) is not visible, and only a small residual contribution of the E$_1$(TO) can be observed, in the form of a component at the low energy side of E$_2^H$. The very low intensity of this E$_1$(TO) is a confirmation of the good crystalline quality of the layers.

The influence of Mg doping on the Raman spectra of GaN:Mg has already been widely discussed in literature~\cite{Brandt1994,Harima1999,Kaschner1999,KirsteJAP2013}. In the studied energy range, only the local vibrational mode (LVM) at 657~cm$^{-1}$ for substitutional Mg in GaN is expected. However, as-grown samples generally do not exhibit this mode, since H is bound to Mg in Mg--N--H complexes, which are Raman-silent in this range~\cite{ManjonJAP2005}. An annealing step is necessary in order to break the complex, to allow H to diffuse towards the sample surface, and to finally reveal the free-Mg LVM~\cite{NakamuraJJAP1992,GotzAPL1996a,PozinaAPL2008}. As expected and as seen in Fig.~\ref{fig:fig1}, the as-grown GaN:Mg considered here does not exhibit the Mg-LVM, confirming that H is bound to Mg. 
 
The introduction of Mn into GaN induces a much richer structure in the spectra and the resulting Raman signature were previously discussed, though the origin of the various contributions is still a matter of controversy~\cite{GuoAPL2006,daSilvaJOP2008,AlarconSST2009}.

\begin{figure}[ht]
	\begin{center}
	\includegraphics[width=0.95\linewidth]{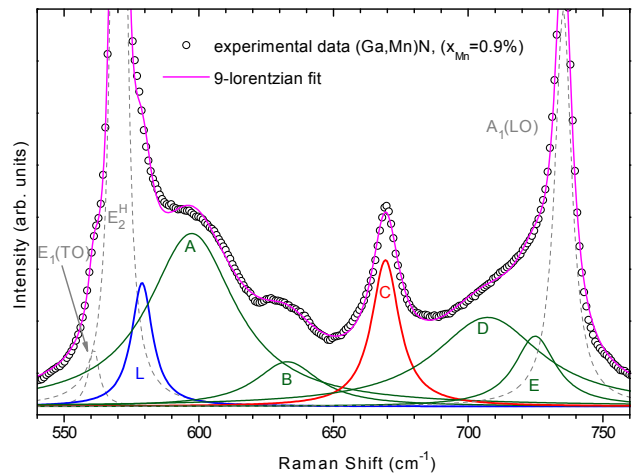}
	\caption{Deconvolution of the Mn-induced Raman broad band in multiple Lorentzians.}
	\label{fig:fig2}
	\end{center}
\end{figure}

To analyse the spectra of (Ga,Mn)N, we employ an approach similar to the one of Gebicki \emph{et al.}~\cite{Gebicki2008}, but based on the deconvolution of the Mn-related signal into six Lorentzian contributions, as shown in Fig.~\ref{fig:fig2}. In addition to the GaN E$_2^H$, E$_1$(TO) and A$_1$(LO), we identify a contribution labeled $L$ which corresponds to the Mn LVM associated to the GaN optical branch. According to the light impurity model~\cite{Gebicki2000, Gebicki2008,Barker1975}, the frequency $\omega^{TO}_{\mathrm{Mn-LVM}}$ of the TO Mn-LVM can be estimated $via$ the formula: 

$$\omega^{TO}_{\mathrm{Mn-LVM}} = \omega^{TO}_{\mathrm{GaN}}\sqrt{\frac{\mu_{GaN}}{\mu_{MnN}}}$$

where $\mu_\mathrm{Ga-N}$ and $\mu_\mathrm{Mn-N}$ correspond to the reduced mass of the Ga--N and M--N pairs, respectively~\cite{Barker1975}.
Considering the frequency of the  E$_2^H$ peak (569~cm$^{-1}$) -- which is the only transverse optical mode allowed in our geometry -- we obtain for the TO Mn-LVM peak a theoretical value of 582~cm$^{-1}$, which is in agreement with the experimental value of 579~cm$^{-1}$. In addition to the Mn-LVM, we can also identify five lorentzian contributions (\emph{A,B,C,D,E}). The integrated intensity of the \emph{C} peak depends linearly on the Mn concentration, and the whole set (\emph{ABCDE}) vanishes when increasing the laser wavelength from 633~nm to 785~nm (not shown), confirming the resonant character of this broad band hinted at in Ref.~\onlinecite{daSilvaJOP2008} and \onlinecite{Gebicki2008}.

\begin{figure}[ht]
	\begin{center}
	\includegraphics[width=0.95\linewidth]{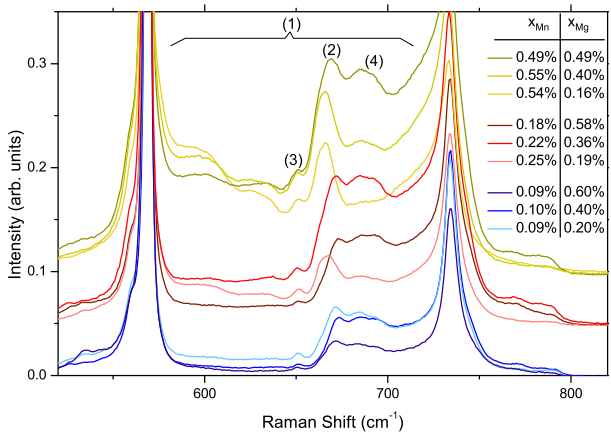}
	\caption{Raman spectra of (Ga,Mn)N:Mg as a function of the Mn and Mg concentrations. The curves are grouped by three: in each group the Mn concentration is kept constant while the Mg concentration is varied. The groups are separated by a $y$-offset of 0.05. The characteristic features of (Ga,Mn)N:Mg are indicated by (1), (2), (3), and (4).}
	\label{fig:fig3}
	\end{center}
\end{figure}

In the samples containing both Mn and Mg, the presence of Mg heavily affects the Mn-related signature in the Raman spectrum. In Fig.\,\ref{fig:fig3}, the Raman spectra for (Ga,Mn)N:Mg samples with different Mn and Mg concentrations ranging from 0.1\% up to 0.5\%, are reported.
One can identify four remarkable features occuring in the presence of both Mn and Mg, namely: (1) a significant quenching of the Mn-induced broad band (\emph{ABDE} and to a lesser extent of \emph{C}); (2) a shift in the position of the \emph{C} peak; (3) the appearance of an additional sharp peak around 650\,cm$^{-1}$; (4) the presence of a specific broad peak around 688\,cm$^{-1}$. In the following, we discuss these new features in the frame of the recently demonstrated evidence of Mn--Mg$_k$ complexes in (Ga.Mn)N:Mg\,\cite{Devillers2012}. In order to provide a quantitative description, we deconvolute the spectra in a similar way as we did in the case of (Ga,Mn)N (without Mg). To take the influence of Mg into account, three Lorentzian contributions are added, one around 650~cm$^{-1}$ to describe feature (3) and two between 680~cm$^{-1}$ and 695~cm$^{-1}$ to describe feature (4). With these two components, it is possible to reproduce the experimental data, and to extrapolate the relevant parameters to describe the evolution of the spectra as a function of Mn and Mg, respectively.

The Mn-induced broad band covering the range between E$_2^H$ and A$_1$(LO) (peaks \emph{A,B,D,E}) is highly dependent on the Mg concentration. The intensity of this broad band is at its maximum in the absence of Mg and is rapidly damped as soon as Mg is added, until it goes under the detection limit for $x_\mathrm{Mg}\geq 2 x_\mathrm{Mn} $. The fact that the attenuation of the \emph{C} peak is in comparison much less evident -- also upon renormalized with the Mn concentration -- is a hint of the fact that, despite the symmetry common to all these peaks as demonstrated by Gebicki \emph{et al.}\,\cite{Gebicki2008}, the physical mechanism behind the origin of the \emph{C} peak is different from the one giving rise to \emph{A,B,D,E}.

\begin{figure}[ht]
	\begin{center}
	\includegraphics[width=0.65\linewidth]{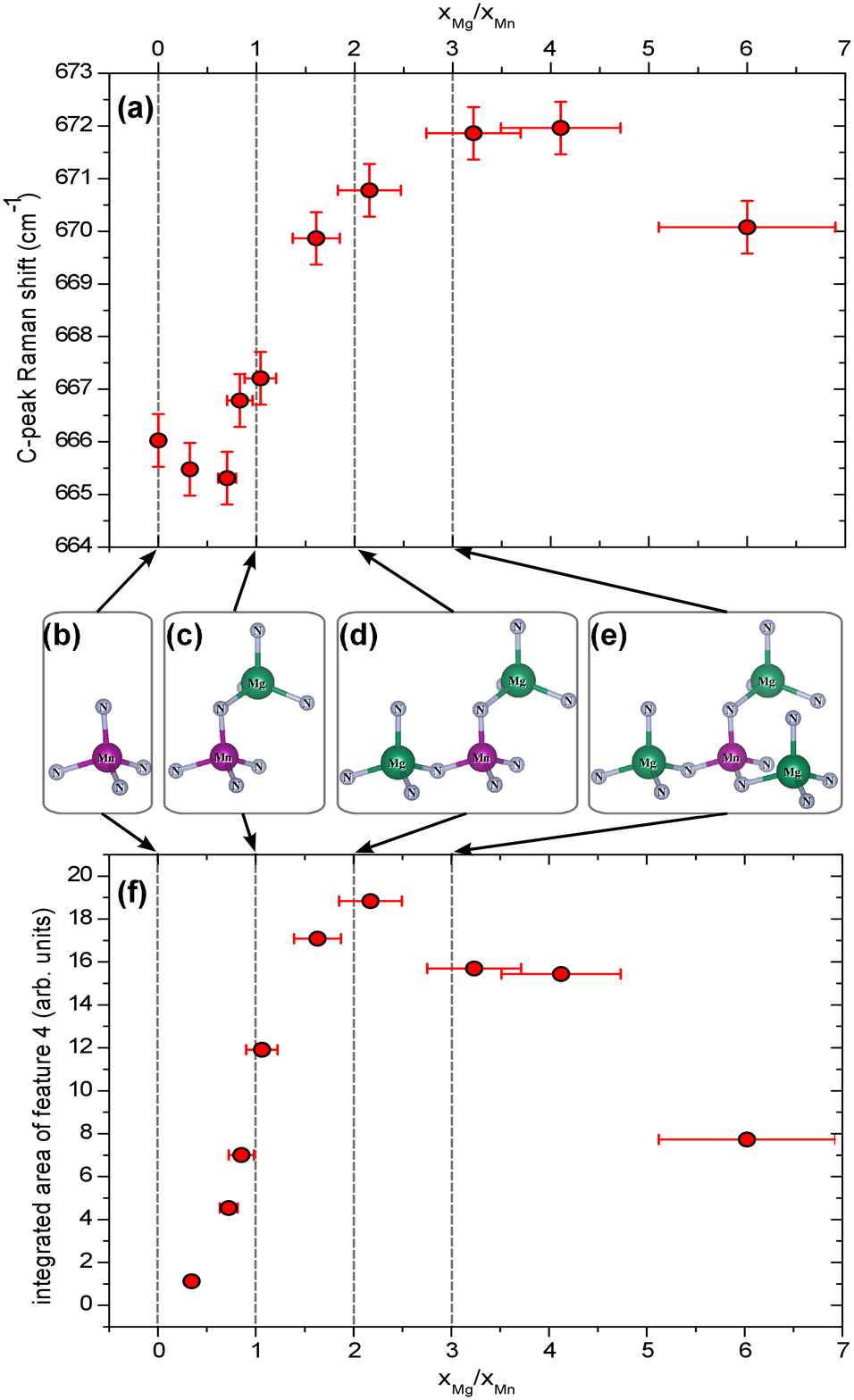}
	\caption{(a) Position of the \emph{C} peak as a function of the $x_\mathrm{Mg}/x_\mathrm{Mn}$ ratio; (b,c,d,e) Structure of Mn--Mg$_k$ complexes for $k=(0,1,2,3)$, corresponding to the ratio $x_\mathrm{Mg}/x_\mathrm{Mn}\,=\,(0,1,2,3)$; (f) Integrated intensity of the Mn--Mg$_k$-related double peak (680~cm$^{-1}$ -- 695~cm$^{-1}$) as a function of the $x_\mathrm{Mg}/x_\mathrm{Mn}$ ratio.}
	\label{fig:fig4}
	\end{center}
\end{figure}

The \emph{C} peak, $i.e.$ the main feature correlated with Mn, and which scales with the Mn concentration, is affected by the Mg incorporation both in energy position and in intensity. We can observe that the intensity variation is mostly due to the variations in the Mn concentration, but the shift in energy is clearly correlated to the introduction of Mg, and does not directly depend on the Mn concentration. The general trend is that the increment of the ratio $x_\mathrm{Mg}/x_\mathrm{Mn}$ between the Mg and Mn concentrations induces a blue shift of the \emph{C} peak. In Fig.\,\ref{fig:fig4}(a) we report the shift of the \emph{C} peak as a function of the ratio between the concentrations of Mg and Mn. Remarkably, for $x_\mathrm{Mg} \geq x_\mathrm{Mn}$, we observe that this shift and the evolution of the Mn--Mg coordination\,\cite{Devillers2012} follow the same trend.
The shift in the \emph{C} peak is correlated to the formation of Mn--Mg$_k$ complexes as the Mg concentration increases. Previous \emph{ab initio} calculations\,\cite{Devillers2012} have evidenced that the incorporation of Mg in the first cation coordination shell of Mn induces a displacement of the N atom binding Mg to Mn towards the Mn atom. The increment in the stretching frequency with the Mg/Mn ratio is completely coherent with the shortening of the Mn--N bonds induced by the presence of Mg. The saturation of this shifts reveals that for a high Mg/Mn ratio, the Mn environment is not perturbed anymore, as shown by synchrotron EXAFS measurements\,\cite{Devillers2012}. However, this simplified model does not account for the red shift of the \emph{C} peak at low $x_\mathrm{Mg}/x_\mathrm{Mn}$. To understand this phenomenon, it is necessary to take into account the spin density calculations reported in Ref.~\onlinecite{Devillers2012}. The simple {Mn--Mg$_{1}$} complex tends to delocalize the spin density on the nitrogen atoms, while the complexes  of higher order ({Mn--Mg$_{2}$},..) reduce this delocalization of the spin density and enhance the electron-phonon coupling. This effect is likely to be responsible for the non-monotonous behaviour of the \emph{C} peak position as a function of $x_\mathrm{Mg}/x_\mathrm{Mn}$.

The peak at 650~cm$^{-1}$ is clearly visible in all samples containing both Mn and Mg. From its energy position, this feature can reasonably be assigned to the Mg local vibration mode (Mg-LVM) which is generally reported at 657~cm$^{-1}$ in activated Mg-doped GaN. However, this LVM is usually not visible in as-grown GaN:Mg, due to the fact that the formation of a Mg--N--H complex significantly distorts the Mg$_\mathrm{Ga}$--N$_4$ tetrahedron\,\cite{Harima1999,Limpijumnong2003}. In our case, the presence of this peak in as-grown samples can be explained taking into account the atomic arrangement of Mg and Mn in GaN. As shown by SIMS measurements\,\cite{Devillers2012}, the presence of Mn in (Ga,Mn)N:Mg induces the H concentration to be between one and two orders of magnitude lower than in GaN:Mg. The incorporation of H is actually hindered by the formation of the Mn--Mg$_k$ complexes, since Mg is electrically passivated by the proximity of a Mn ion. On the other hand, the presence of Mn in the vicinity of Mg shifts N towards Mn, increasing in this way the Mg--N bond length from 2.03~\AA$ $  to 2.11~\AA. This distortion of the Mg--N$_4$ tetrahedron is much smaller than the one induced by the formation of a Mg--N--H complex \cite{Limpijumnong2003} and does not impair the Mg-LVM. However the red-shift observed (peak maximum at 650~cm$^{-1}$ instead than at 657~cm$^{-1}$) is likely to originate from this distortion. This red-shift does not depend on the Mg concentration, in accordance with the fact that the environment of the Mg atom does not depend massively on the Mn and Mg concentrations (independently of the Mn and Mg concentrations, each Mg is bound to one and only one Mn).

The last feature induced by the presence of Mg in (Ga,Mn)N:Mg is a relatively broad band around 688~cm$^{-1}$, which -- particularly at high $x_\mathrm{Mg}/x_\mathrm{Mn}$ -- is actually composed of at least two peaks, between 680~cm$^{-1}$ and 695~cm$^{-1}$. To describe phenomenologically this mode, we report in Fig.\,\ref{fig:fig4}(f) the integrated intensity of this band (the sum of the integrated intensity of the two peaks), normalized by the Mn concentration, as a function of the ratio between the Mg and Mn concentrations. If we compare the trend in Fig.~\ref{fig:fig4}(f) to the populations of the different complexes calculated in Ref.~\onlinecite{Devillers2012}, it appears that all three complexes give a contribution, and particularly the Mn--Mg$_2$ complex. Surprisingly, according to previous \emph{ab initio} calculations, this complex is the one with the lowest symmetry, the Mn--N$_4$ tetrahedron being more distorted than for other complexes. Since a purely structural argument cannot account for the Raman activity of the Mn--Mg$_2$ complex, one should also consider the influence of the electronic structure. In addition to differences in the localization of carriers for different complexes that we already mentionned, the presence of free Mg in GaN for $x_\mathrm{Mg}/x_\mathrm{Mn}\,\geq\,3$ and the consequent introduction of holes is also expected to affect the Raman activity of this Mn-Mg$_k$ -induced Raman band, and to be responsible for the quenching at high $x_\mathrm{Mg}/x_\mathrm{Mn}$.

In conclusion, Raman spectroscopy measurements at room temperature of (Ga,Mn)N:Mg have shown that the codoping of GaN with Mn and Mg generates Mn--Mg$_k$ cation complexes, in agreement with previously reported synchrotron EXAFS measurements. Moreover, the evolution of the different features relative to Mg and Mn in (Ga,Mn)N:Mg supports the \emph{ab initio} calculations of the local arrangements around the complexes, in particular the displacement of the N atoms surrounding Mn. This new information, contributes to the understanding and control of the functional complexes Mn--Mg$_k$ that have been shown to introduce into the system two quite spectacular features, namely: (i) a broadband infrared luminescence persisting at room temperature and (ii) the possibility to tune the spin state of Mn in a controlled way as a function of the degree of coordination $k$, paving the way to the realization of novel optically active and tunable spintronic devices.\\

This work was supported by the FunDMS Advanced Grant of the European Research Council (ERC Grant No. 227690) within the Ideas
7th Framework Programme of the European Community, by the Austrian Science Fundation -- FWF (P22471, P20065 and P22477) and by the  Fapesp (Grants 2006/05627-8 and 2012/21147-7)

%\bibliography{Devillers-APL-Raman}
%\bibliographystyle{apsrev}

\end{document}